\documentclass[aps, pre, reprint,superscriptaddress,amsmath,amssymb,aps,nofootinbib]{revtex4-2}
\usepackage{graphicx}
\usepackage[colorlinks,linkcolor=blue,citecolor=blue,urlcolor = blue]{hyperref}
\usepackage{mathtools}

\begin{document}

\title{Partitioning Law of Polymer Chains into Flexible Polymer Networks} 

\author{Haruki~Takarai}
\thanks{These authors contributed equally: H. Takarai, T. Yasuda}
\affiliation{Department of Chemistry and Biotechnology, The University of Tokyo, Hongo, Bunkyo-ku, Tokyo 113-8656, Japan}
\author{Takashi~Yasuda}
\thanks{These authors contributed equally: H. Takarai, T. Yasuda}
\affiliation{Faculty of Advanced Life Science, Hokkaido University, Sapporo, Hokkaido 001-0021, Japan}
\author{Naoyuki~Sakumichi}
\email[Corresponding author. ]{sakumichi@gel.t.u-tokyo.ac.jp}
\affiliation{Department of Chemistry and Biotechnology, The University of Tokyo, Hongo, Bunkyo-ku, Tokyo 113-8656, Japan}
\affiliation{Faculty of Social Informatics, ZEN University, Shinjuku, Zushi, Kanagawa 249-0007, Japan}
\author{Takamasa~Sakai}
\email[Corresponding author. ]{sakai@gel.t.u-tokyo.ac.jp}
\affiliation{Department of Chemistry and Biotechnology, The University of Tokyo, Hongo, Bunkyo-ku, Tokyo 113-8656, Japan}

\date{\today}

\begin{abstract}
The equilibrium partitioning of linear polymer chains into flexible polymer networks is governed by intricate entropic constraints arising from the configurational degrees of freedom of both chains and networks; however, a quantitative understanding remains elusive.
Using model hydrogels with precisely defined network structures, we experimentally demonstrate a universal law governing the partitioning of linear polymers into flexible polymer networks.
We establish a label-free contactless method to measure the partition ratio, based on the increase in the osmotic pressure induced by the partitioning of the external polymer chains.
Moreover, we reveal a universal law in which the partition ratio is determined solely by $R_g / l_\mathrm{cycle}$, where $R_g$ is the gyration radius of the polymer chain and $l_\mathrm{cycle}\equiv \xi^{-1/3}$ is the characteristic mesh size of the network, defined by the cycle rank $\xi$, i.e., the number density of elastically effective cycles.
\end{abstract}

\maketitle
\section{Introduction}
Solute partitioning in polymer gels is governed by complex physical and chemical interactions between solutes and the polymer network.
Polymer gels are a unique class of materials in which a solvent-swollen polymer network facilitates solute exchange with the external environment. 
The equilibrium concentration ratio of the solutes between the gel ($\phi_\mathrm{int}$) and an external solution ($\phi_\mathrm{ext}$), defined as the partition ratio $K\equiv \phi_\mathrm{int}/\phi_\mathrm{ext}$, depends on the interactions between the solutes and the polymer network~\cite{ogston1958Spaces, laurent1964Theory, schnitzer1988Analysis, williams1998Partition, hagel2013Diffusion, kotsmar2012Aqueous, tong1996Partitioning, richbourg2023Solute, kremer1994PoreSize, johnson1995Diffusion, tsonchev2007Partitioning, sleeboom2017Compression, gehrke1997Factors, horkay1986Studies}.
Typically, $K$ decreases with increasing solute size, similar to behavior in rigid, static porous media \cite{teraoka1996Polymer, casassa1967Equilibrium, giddings1968Statistical, doi1975Equilibrium}.
Such size-dependent partitioning plays a critical role in applications ranging from membrane separation and size-exclusion chromatography \cite{gorbunov1988Fundamentals} to regulation of solute distribution in biological systems such as membrane-bound organelles and biomolecular condensates \cite{sugano2010Coexistence, wei2017Phase}.

Despite its importance, the microscopic mechanism governing network-dependent variations in $K$ has remained elusive.
According to the Nernst distribution law~\cite{berthelot1872, nernst1891Verteilung}, which is a fundamental principle in physical chemistry~\cite{atkins2013Elements}, the partition ratio should be concentration-independent in the dilute regime, allowing for the definition of a constant partition ratio $K_0$.
Descriptions of $K_0$ often rely on obstacle models such as the Ogston model~\cite{ogston1958Spaces, laurent1964Theory, williams1998Partition, hagel2013Diffusion, schnitzer1988Analysis, kotsmar2012Aqueous, tong1996Partitioning, richbourg2023Solute, johnson1995Diffusion}.
The Ogston model characterizes the network as a set of rigid rods; however, the rod size depends on network concentration and does not correspond to the actual fiber radius of polymer chains~\cite{williams1998Partition, hagel2013Diffusion}.
Moreover, experimentally measured $K$ frequently varies with solute concentrations in gels \cite{schnitzer1988Analysis}, contradicting the Nernst distribution law, which predicts constant $K_0$.
A thermodynamic approach~\cite{horkay1986Studies} based on osmotic equilibrium has also been applied to analyze $K$; however, it has not provided a quantitative explanation.

These theoretical inconsistencies reflect the experimental challenges involved in accurate measurements of $K$.
First, polymer gels typically exhibit significant structural and spatial inhomogeneity \cite{shibayama1998Spatial}, resulting in nonuniform solute distributions. 
Second, measurements using fluorescently labeled solutes \cite{
williams1998Partition, hagel2013Diffusion, kotsmar2012Aqueous, tong1996Partitioning, richbourg2023Solute, johnson1995Diffusion} are susceptible to unintended chemical interactions and molecular aggregation, leading to systematic deviations in the measured values, particularly under conditions far from the ideal dilute regime. 
Moreover, these factors may interact synergistically, further complicating the accurate determination of partition ratios.

\begin{figure*}[t!]
\centering
\includegraphics[width=\linewidth]{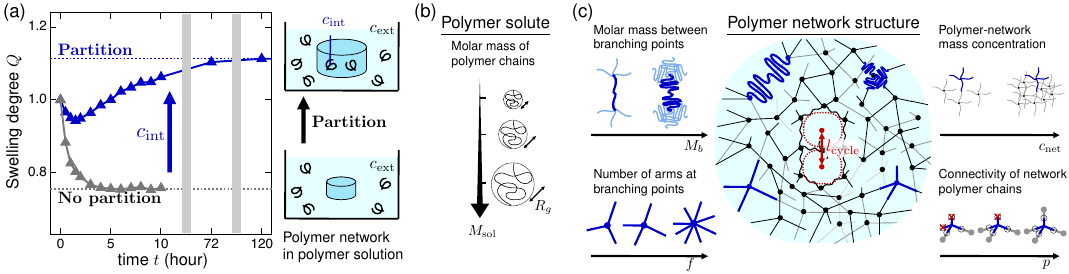}
\caption{Evaluation of partition ratio $K$ of polymer chains into polymer gels with controlled network structures.
(a) Time evolution of the swelling ratio in polymer gels immersed in external polymer solutions with concentration $c_\mathrm{ext}$, induced by increasing the internal concentration of partitioned polymer chains $c_\mathrm{int}$.
When external polymer chains do not partition into the gel ($c_\mathrm{int} \approx 0$), the gel deswells to reach equilibrium (gray dashed line).
In contrast, when external polymer chains partition into the gel ($c_\mathrm{int}>0$), the osmotic pressure increases with $c_{\mathrm{int}}$ [Eq.~(\ref{eq:Pimixcin})], causing the gel to swell further until reaching equilibrium (blue dashed line).
(b) Control of solute size by varying the molar mass of polymer chains $M_\mathrm{sol}$ to have different gyration radii $R_g$.
(c) Polymer network structures with different network cycle lengths $l_\mathrm{cycle}$ determined by the molar mass between branching points $M_b$, polymer-network mass concentration $c_\mathrm{net}$, number of arms at branching points $f$, and network connectivity $p$.
}
\label{fig:abst}
\end{figure*}

In this article, we develop an unlabeled, contactless method for measuring the partition ratio $K$ to overcome these experimental challenges.
Using the physical laws governing osmotic pressure in polymer gels \cite{flory1953Principles, horkay1986Studies, yasuda2020Universal, sakumichi2022Semidilute}, we establish a scaling relationship between osmotic pressure and total polymer concentration, enabling accurate determination of the partitioned polymer concentration through measurements of the equilibrium swelling ratio [Fig.~\ref{fig:abst}(a)]. 
Model gel systems formed from star-shaped polymers \cite{sakai2008design} provide an ideal platform for such measurements \cite{li2013Migration, li2014Electrophoretic, fujiyabu2017Permeation, fujiyabu2019Diffusion} because they permit precise tuning of the network structure.
Here, the use of chemically identical PEG for both the solute and the network eliminates specific interactions (e.g., $\chi$-parameter effects or preferential adsorption), allowing us to isolate the purely geometrical and topological constraints governing partitioning.

We systematically investigate the partitioning behavior of polymer chains in model gels across a broad range of polymer solute sizes and network structures [Fig.~\ref{fig:abst}(b) and (c)].
Our findings demonstrate that the Nernst distribution law holds for model gels in the dilute regime, with the experimentally measured partition ratio $K$ becoming constant ($K = K_0$).
Furthermore, by examining the characteristic length scales (mesh sizes) of the polymer network for various densities and topologies, we reveal that $K_0$ is governed by a universal law based on the ratio of the solute gyration radius ($R_g$) to the network cycle length ($l_\mathrm{cycle}$), deviating significantly from the predictions of the Ogston model.

\section{Theory of Equilibrium Swelling in Polymer Gels}
We consider the equilibrium swelling of polymer gels immersed in external solutions, both with and without solute partitioning.
First, we review the osmotic equilibrium of polymer gels in external polymer solutions, assuming negligible partitioning into the gel~\cite{flory1953Principles, horkay1986Studies, yasuda2020Universal, sakumichi2022Semidilute}.
Throughout this study, we adopt the polymer mass concentration $c$, defined as the polymer mass per unit volume of solvent \cite{yasuda2020Universal}, to describe the equilibrium swelling of polymer gels.
The volume fraction $\phi$ is given by $\phi=c/(\rho_p + c)$, where $\rho_p$ is the mass density of the polymer.
When a polymer gel is immersed in an external solution with mass concentration $c_\mathrm{ext}$ that exerts osmotic pressure $\Pi_\mathrm{ext}$, the gel reaches equilibrium by adjusting its weight ($W_0 \to W_\mathrm{eq}$) and polymer-network mass concentration ($c_\mathrm{net,0} \to c_\mathrm{net}$).
Osmotic equilibrium occurs when the internal osmotic pressure, composed of the polymer-solvent mixing contribution ($\Pi_\mathrm{mix}$) and the network elastic contribution ($\Pi_\mathrm{el}$), balances the external osmotic pressure ($\Pi_\mathrm{ext}$):
\begin{equation}
\Pi_\mathrm{mix} + \Pi_\mathrm{el} = \Pi_\mathrm{ext}.
\label{eq:equilibrium}
\end{equation}
For gels with sufficient polymer-solvent miscibility, $\Pi_\mathrm{mix}$ follows the semidilute scaling law, $\Pi_\mathrm{mix} \propto c_{\mathrm{net}}^{3\nu/(3\nu-1)}$ \cite{yasuda2020Universal, sakumichi2022Semidilute, des1975lagrangian, de1979Scaling}, where $\nu \approx 0.588$ is the critical exponent for self-avoiding walks in three dimensions~\cite{flory1953Principles,pelissetto2002critical}.
The prefactor depends on the polymer-solvent interactions and temperature. 
The elastic contribution is given by $\Pi_\mathrm{el} = -G_0(c_\mathrm{net}/c_\mathrm{net,0})^{1/3}$ \cite{james1949simple, horkay2000osmotic}, where $G_0$ is the shear modulus of the gel in its as-prepared state.
Using Eq.~\eqref{eq:equilibrium}, when $c_\mathrm{net,0}$, $G_0$, and the prefactor of $\Pi_\mathrm{mix}$ are known, we can determine equilibrium $c_\mathrm{net}$ from $\Pi_\mathrm{ext}$ without direct measurements.
Alternatively, $c_\mathrm{net}$ can be calculated directly from the equilibrium mass swelling ratio defined as
$Q_\mathrm{eq} = W_\mathrm{eq}/W_0 = 
(c_\mathrm{net,0}/c_\mathrm{net})
(\rho_s + c_\mathrm{net}) / 
(\rho_s + c_\mathrm{net,0})$, 
where $\rho_s$ is the mass density of a pure solvent.
In this study, we use $\rho_p \approx 1129$ kg$/$m${}^3$ and $\rho_s \approx 997$ kg$/$m${}^3$ for polymers [poly(ethylene glycol)] and aqueous solvents~\cite{kell1975Density}, respectively.

Next, we extend the osmotic equilibrium described by Eq.~(\ref{eq:equilibrium}) to systems in which the external polymer solutes partition into the polymer gel.
In this case, the polymer solutes entering the gel increase $\Pi_\mathrm{mix}$.
When the solute consists of polymer chains chemically identical to those forming the network, both the network polymer (with mass concentration $c_\mathrm{net}$) and the partitioned polymer chains (with mass concentration $c_\mathrm{int}$) contribute to the mixing term as
\begin{equation}
\Pi_{\mathrm{mix}}^{} =A (c_{\mathrm{net}}+{c_{\mathrm{int}}})^{3\nu/(3\nu-1)},
\label{eq:Pimixcin}
\end{equation}
where the total polymer mass concentration $c = c_\mathrm{net} + c_\mathrm{int}$ determines $\Pi_\mathrm{mix}$ using the scaling law applied in the absence of partitioning.
Equation~(\ref{eq:Pimixcin}) indicates that regardless of their origin (network or partitioned chains), the polymer segments contribute equally to $\Pi_\mathrm{mix}$.

Below, we experimentally validate Eq.~(\ref{eq:Pimixcin}), rather than an (incorrect) alternative additive decomposition:
\begin{equation}
\Pi_{\mathrm{mix}}^{} = \Pi_{\mathrm{mix}}^{\mathrm{(net)}} + \Pi_{\mathrm{mix}}^{\mathrm{(int)}},
\label{eq:Piint}
\end{equation}
where $\Pi_{\mathrm{mix}}^{\mathrm{(net)}} \propto  c_{\mathrm{net}}^{3\nu/(3\nu-1)}$ and 
$\Pi_{\mathrm{mix}}^{\mathrm{(int)}} \propto c_{\mathrm{int}}^{3\nu/(3\nu-1)}$ are the mixing contributions from the network and partitioned chains, respectively.
Our results demonstrate that Eq.~(\ref{eq:Pimixcin}) provides the correct description, confirming that osmotic pressure arises from the collective interactions of all polymer segments rather than from the separate contributions of distinct polymer populations.

\section{Materials and Methods}
We used polymer gels with precisely controllable network structures \cite{sakai2008design} as model systems to investigate the network partitioning law.
We synthesized star polymer networks through the end-to-end polymerization of two types of star polymers with mutually reactive terminal functional groups, enabling precise control of the molar mass between branching points ($M_b$), network connectivity ($p$), branching number ($f$), and polymer-network mass concentration ($c_\mathrm{net,0}$).
We dissolved each star polymer [poly(ethylene glycol), PEG] with either maleimide (MA) or thiol (SH) terminal functional groups (XIAMEN SINOPEG BIOTECH Co., Ltd., China) in an aqueous citrate-phosphate buffer at \textit{p}H $3.4$--$3.8$ with an ionic strength of $0.1$ M for $c_\mathrm{net,0} = 30, 60$, and $90$~g$/$L. 
For gelation, we mixed these two solutions with equal molar masses of the precursors $M_\mathrm{pre} = 10, 20,$ and $40$ kg$/$mol and identical arm numbers ($f = 3$, $4$, and $8$) at a volume mixing fraction $s$ ($0 \leq s \leq 1/2$).
This allowed systematic variation of $M_b=2M_\mathrm{pre}/f$ and $p = 2s$ \cite{yoshikawa2019connectivity}, where $p$ is the fraction of reacted to total terminal functional groups ($0 \leq p \leq 1$).
We poured each mixed solution into a disk-shaped mold ($10$ mm diameter, $2$ mm height), and maintained the samples at a temperature of $T = 298$~K for $1$--$3$~days to ensure reaction completion.

We immersed each gel sample in an excess external aqueous linear-PEG solution with a controlled concentration $c_\mathrm{ext}$ and osmotic pressure $\Pi_\mathrm{ext}$.
We prepared these external PEG solutions with molar masses $M_\mathrm{sol}=5, 10,$ and $20$~kg$/$mol (XIAMEN SINOPEG BIOTECH Co., Ltd., China), exhibiting narrow molar mass distributions $M_w/M_n < 1.05$, for $c_\mathrm{ext} = 10$--$120$~g$/$L dissolved in deionized water.
We determined $\Pi_\mathrm{ext}$ using the established osmotic equation of state for linear-PEG solutions \cite{noda1981thermodynamic,higo1983osmotic,yasuda2023Universality} (detailed in the Supplemental Material (SM), Sec.~S1).
After confirming equilibrium (see SM,~Sec.~S2), we measured the weight change of each gel sample ($W_0 \to W_\mathrm{eq}$) using an electronic scale.
We verified that buffer salts initially present in the gel do not measurably affect partitioning and swelling in linear-PEG solutions (detailed in SM, Sec.~S3).

\begin{figure}[t!]
\centering
\includegraphics[width=\linewidth]{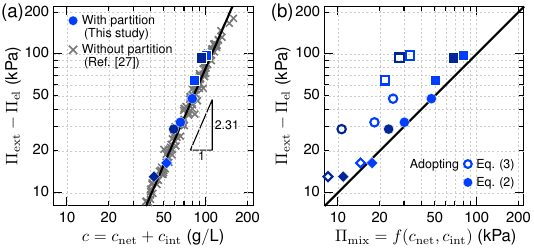}
\caption{Experimental verification of $\Pi_\mathrm{mix}$ in polymer gels containing partitioned polymer chains.
(a) Semidilute scaling law [Eq.~(\ref{eq:Pimixcin}), black solid line] governing $\Pi_\mathrm{mix}=\Pi_\mathrm{ext}-\Pi_\mathrm{el}$ in polymer gels with (blue filled symbols) and without 
(gray crosses) partitioned chains~\cite{yasuda2020Universal, sakumichi2022Semidilute}.
(b) Comparison between measured $\Pi_\mathrm{mix}=\Pi_\mathrm{ext}-\Pi_\mathrm{el}$ and calculated $\Pi_\mathrm{mix}$ using Eqs.~(\ref{eq:Pimixcin}) and (\ref{eq:Piint}) (filled and open symbols, respectively).
}
\label{fig:pimix}
\end{figure}

\section{Osmotic pressure in gels containing partitioned polymer chains}
To verify Eq.~(\ref{eq:Pimixcin}), we directly quantified $c_\mathrm{int}$ by drying the gels to remove the solvent and measuring the dry weight $W_\mathrm{dry} = W_\mathrm{net} + W_\mathrm{int}$, where $W_\mathrm{net}$ and $W_\mathrm{int}$ are the weights of the network and partitioned chains, respectively.
We prepared cylindrical gel samples ($5.3$ mm diameter, $4.8$ mm height), equilibrated them in external PEG solutions, and vacuum-dried them at $T = 333$~K for at least $6$~hours.
We used a thermogravimetric analyzer (DTG-60H, SHIMADZU, Japan) to accurately measure $W_\mathrm{dry}$ (see SM,~Sec.~S4 for detailed results). 
Because $W_\mathrm{net} = W_0[c_\mathrm{net,0}/ (\rho_s + c_\mathrm{net,0})]$ remained constant during swelling and deswelling, we calculated 
$c_\mathrm{net} = \rho_s W_\mathrm{net} / (W_\mathrm{eq} - W_\mathrm{dry})$ and $c_\mathrm{int} = \rho_s (W_\mathrm{dry} - W_\mathrm{net})/(W_\mathrm{eq} - W_\mathrm{dry})$.

We demonstrate that Eq.~(\ref{eq:Pimixcin}), rather than Eq.~(\ref{eq:Piint}), governs $\Pi_\mathrm{mix}$ in polymer gels containing partitioned polymer chains.
By directly measuring the internal polymer concentration ($c_\mathrm{int}$) via drying, we calculate $\Pi_\mathrm{mix} = \Pi_\mathrm{ext} - \Pi_\mathrm{el}$ for the total polymer concentration $c=c_\mathrm{net} + c_\mathrm{int}$.
To evaluate $\Pi_\mathrm{el}$, we used the previously measured shear modulus $G_0$~\cite{yoshikawa2021Negative} (detailed in SM, Sec.~S5).
As shown in Fig.~\ref{fig:pimix}(a), Eq.~(\ref{eq:Pimixcin}) (black solid line) accurately describes the measured $\Pi_\mathrm{mix}$ in gels containing partitioned polymer chains (blue filled symbols).
The prefactor $A \approx 1.9$ was obtained from these plots and agrees with the value previously reported for gels without partitioning (gray crosses) in aqueous PEG systems at $T=298$~K~\cite{yasuda2020Universal, sakumichi2022Semidilute}.
By contrast, Fig.~\ref{fig:pimix}(b) reveals that the experimentally determined $\Pi_\mathrm{mix}$ deviates significantly from predictions using Eq.~(\ref{eq:Piint}) (open symbols), whereas calculations using Eq.~(\ref{eq:Pimixcin}) (filled symbols) agree well with experiment. 
This agreement enables the evaluation of $c_\mathrm{int}$ using Eqs.~(\ref{eq:equilibrium}) and (\ref{eq:Pimixcin}).

\begin{figure}[t!]
\centering
\includegraphics[width=\linewidth]{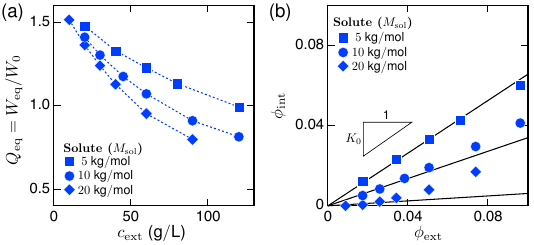}
\caption{Evaluation of constant partition ratio $K_0$ demonstrating the Nernst distribution law ($K \approx K_0$).
Typical results for $M_\mathrm{sol} = 5, 10,$ and $20$ kg$/$mol (squares, circles, and diamonds, respectively) partitioning into polymer networks with identical conditions: $M_\mathrm{pre} = 20$ kg$/$mol for $c_{\mathrm{net,0}} = 60$~g$/$L at $s=0.5$ and $f=4$. 
(a) Dependence of equilibrium swelling ratio $Q_\mathrm{eq} = W_\mathrm{eq}/W_0$ on $c_\mathrm{ext}$.
As $c_{\mathrm{ext}}$ increases, $\Pi_\mathrm{ext}$ increases and $Q_\mathrm{eq}$ decreases according to Eq.~(\ref{eq:equilibrium}).
(b) Dependence of $\phi_\mathrm{int}$ on $\phi_\mathrm{ext}$, where $\phi_{\mathrm{int}}$ are evaluated using Eqs.~(\ref{eq:equilibrium}) and (\ref{eq:Pimixcin}).
We determine the constant partition ratio $K_0 = \phi_\mathrm{int}/\phi_\mathrm{ext}$ (black solid line) using data from the dilute-concentration regime ($\phi_\mathrm{ext}/\phi_\mathrm{ext}^{*} < 1$), where partitioned linear polymers are isolated; see SM, Sec.~S1 for $\phi_\mathrm{ext}^{*}$.
}
\label{fig:result1}
\end{figure}

\section{Nernst distribution law for polymer chain partitioning into polymer networks}
We systematically measured the equilibrium swelling ratio $Q_\mathrm{eq}$ for various molar masses of the polymer solute ($M_\mathrm{sol}$) and network structures characterized by $M_b$, $c_\mathrm{net,0}$, $p$, and $f$.
Our results demonstrate that the Nernst distribution law ($K \approx K_0$) holds for polymer chain partitioning in polymer networks.
As shown in Fig.~\ref{fig:result1}(a), $Q_\mathrm{eq}$ decreases with increasing $c_\mathrm{ext}$ and $M_\mathrm{sol}$ (all other data are presented in SM, Sec.~S2). 
Using Eqs.~(\ref{eq:equilibrium}) and (\ref{eq:Pimixcin}), we evaluated $\phi_\mathrm{int}$.
Figure~\ref{fig:result1}(b) shows that $\phi_\mathrm{int}$ is directly proportional to $\phi_\mathrm{ext}$ in the dilute-concentration regime (black solid lines), indicating a constant partition ratio $K_0 = \phi_\mathrm{int}/\phi_\mathrm{ext}$, which is consistent with the Nernst distribution law \cite{nernst1891Verteilung}.
Notably, when $\phi_\mathrm{ext}$ exceeds a critical threshold ($\phi_\mathrm{ext}/\phi_\mathrm{ext}^{*} > 1$), this relationship breaks down as higher-order concentration terms become significant ($K > K_0$), with $\phi_\mathrm{ext}^{*}$ decreasing as $M_\mathrm{sol}$ increases (detailed in SM, Sec.~S1).

\begin{figure}[t!]
\centering
\includegraphics[width=0.9\linewidth]{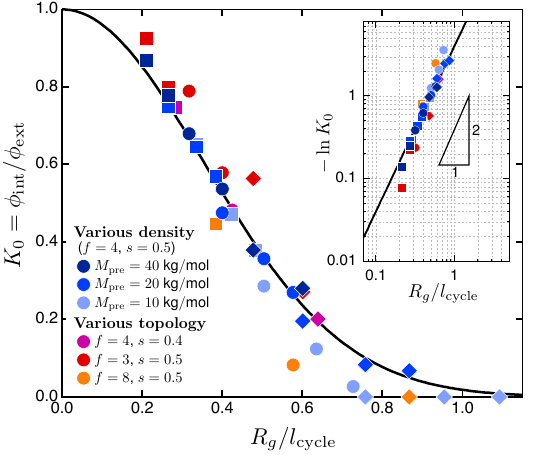}
\caption{Universal law governing constant partition ratio $K_0$ of polymer chains in polymer networks. 
We systematically varied polymer network structures by tuning network density with $M_\mathrm{pre}\, (= fM_b/2) = 10$, $20$, and $40$~kg$/$mol for $c_\mathrm{net,0} = 30$, $60$, and $90$~g$/$L, network topology with $f=3$, $4$, and $8$ (red, blue, and orange symbols) at $p=0.8$ and $1.0$ (purple and blue symbols), and polymer solute size with $M_\mathrm{sol}=5$, $10$, and $20$~kg/mol (squares, circles, and diamonds).
All $K_0$ values follow Eq.~(\ref{eq:K}) (black solid curve) as a function of the ratio between the gyration radius of the polymer solute and the cycle length $R_g/l_\mathrm{cycle}$. 
Inset: Log-log plot showing $-\ln K_0 \propto (R_g/l_\mathrm{cycle})^2$ [Eq.~(\ref{eq:K})], except for $K_0=0$.
}
\label{fig:mainresult}
\end{figure}

\section{Universal law of constant partition ratio for polymer chain partitioning into polymer networks}
We reveal a universal law governing $K_0$ for polymer chain partitioning into polymer networks.
To identify the characteristic length scale governing partitioning, we examine three candidates: the mean distance between branching points $l_{\mathrm{branch}}\,(=n^{-1/3})$, between elastically effective junction points $l_{\mathrm{junction}}\,(=\mu^{-1/3})$, and between the centroids of elastically effective cycles $l_{\mathrm{cycle}}\,(=\xi^{-1/3})$.
Here, $n$, $\mu$, and $\xi$ are the number densities of total precursors, elastically effective junction points, and elastically effective cycles (cycle rank), respectively, calculated via the Bethe approximation~\cite{macosko1976new, miller1976new,yoshikawa2019connectivity} (see Appendix B).

Figure~\ref{fig:mainresult} shows the measured $K_0$ values plotted against the dimensionless ratio $R_g/l_\mathrm{cycle}$.
We find that the universal partitioning law
\begin{equation}
K_0 = \exp \left[ -4 \left({R_g} / {l_\mathrm{cycle}} \right)^2 \right]
\label{eq:K}
\end{equation}
accurately describes all data over a wide range of polymer solute sizes and network structures. 
By contrast, $l_{\mathrm{branch}}$ and $l_{\mathrm{junction}}$ fail to collapse the data for networks with varying $f$ and $p$, except for the data with $f=4$ and $p=1$, for which $l_{\mathrm{branch}} = l_{\mathrm{junction}} = l_{\mathrm{cycle}}$ (see Fig.~\ref{fig:meshsize} in Appendix~C).
Using the functional form $K_0 = \exp[-4 (R_g / l)^2]$ with $l = l_{\mathrm{branch}}$, $l_{\mathrm{junction}}$, or $l_{\mathrm{cycle}}$, the root mean squared deviations between measured and predicted $K_0$ for the remaining data are $0.24$, $0.22$, and less than $0.1$, respectively, confirming that the cycle length is the governing parameter (Fig.~\ref{fig:meshsize}).

We also compared our results with the conventional Ogston model~\cite{ogston1958Spaces, laurent1964Theory, williams1998Partition, hagel2013Diffusion, schnitzer1988Analysis, kotsmar2012Aqueous, tong1996Partitioning, richbourg2023Solute, johnson1995Diffusion}, which predicts $K_0 = e^{-\phi_\mathrm{excl}}$ based on the excluded volume fraction $\phi_\mathrm{excl}$ of the network for the solute.
The measured $K_0$ values systematically exceed the predictions of the Ogston model (see Fig.~\ref{fig:Ogston} in Appendix~A).
This discrepancy arises because the Ogston model treats polymer networks as rigid rod-like obstacles, thereby neglecting network topology.

\begin{figure}[t!]
\centering
\includegraphics[width=\linewidth]{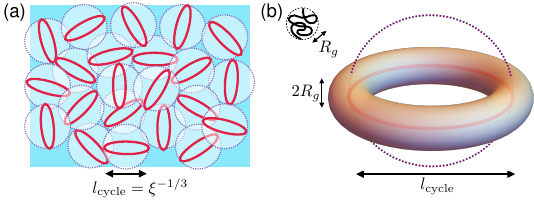}
\caption{Cycle-based model for the partitioning of polymer chains into polymer networks.
(a) Schematic of a polymer network composed of elastically effective cycles (red circles), each associated with a representative volume element of size $\sim l_\mathrm{cycle}^3$ (blue dashed circles).
(b) Steric exclusion around a single cycle: the center of a guest polymer chain with gyration radius $R_g$ is excluded from a torus-like region (gray shaded) with tube radius $R_g$ surrounding the cycle within a volume element of size $\sim l_\mathrm{cycle}^3$.
}
\label{fig:model}
\end{figure}

To rationalize the universal law in Eq.~(\ref{eq:K}), we propose a cycle-based model [Fig.~\ref{fig:model}].
We consider a volume element of size $\sim l_\mathrm{cycle}^3$ associated with a single elastically effective cycle in the network [Fig.~\ref{fig:model}(a)].
The center of a guest chain (i.e., a partitioned polymer chain) with gyration radius $R_g$ is sterically excluded from a torus-like region with tube radius $R_g$ and volume $\sim R_g^2 l_\mathrm{cycle}$ surrounding the cycle [Fig.~\ref{fig:model}(b)].
According to the Poisson distribution, the probability of partitioning corresponds to the zero-encounter probability, $P(0) = e^{-\langle n\rangle}$, where the mean number of encounters with obstructing cycles is $\langle n\rangle \propto R_g^2/l_\mathrm{cycle}^2$.
Setting $K_0=P(0)$~\cite{ogston1958Spaces} yields $\ln K_0 \sim -(R_g/l_\mathrm{cycle})^2$, recovering the functional form of Eq.~(\ref{eq:K}).
In the dilute regime, the chemical-potential difference of the solute between the gel and the external solution ($\Delta\mu$) determines the partition ratio via $K_0=\exp(-\Delta\mu/k_BT)$.
Our model shows $\Delta\mu \propto k_BT(R_g/l_\mathrm{cycle})^2$ from an entropic confinement penalty imposed by the network topology, thereby extending the thermodynamic treatment~\cite{horkay1986Studies} to explicitly include topological constraints.
This framework is analogous to the Casassa model for partitioning in rigid porous media~\cite{casassa1967Equilibrium, kremer1994PoreSize}; in our model, elastically effective cycles impose conformational constraints on guest chains.
Thus, elastically effective cycles serve as the fundamental structural units governing partitioning in polymer networks.

\section{Concluding remarks}
We experimentally investigated the partitioning of polymer chains into polymer networks for various solute sizes and network structures (Fig.~\ref{fig:abst}).
By extending the semidilute scaling law [Eq.~(\ref{eq:Pimixcin})] to gels containing partitioned chains (Fig.~\ref{fig:pimix}), we systematically measured the partition ratio $K$ using osmotic equilibrium and demonstrated that the Nernst distribution law holds with a constant $K_0$ (Fig.~\ref{fig:result1}).
The obtained $K_0$ is inconsistent with the predictions of the conventional Ogston model (Fig.~\ref{fig:Ogston}) and instead follows the universal partitioning law [Eq.~(\ref{eq:K})] (Fig.~\ref{fig:mainresult}).

These findings establish $l_\mathrm{cycle}$ as the characteristic length scale governing partitioning in gel networks (Figs.~\ref{fig:mainresult} and \ref{fig:meshsize}).
Notably, the same length scale ($l_\mathrm{cycle}$) also governs scaling properties of elastic modulus~\cite{yoshikawa2019connectivity,lang2019elasticity,yoshikawa2021Negative,sakumichi2021linear}, fracture behavior~\cite{masubuchi2023Phantom,masubuchi2024Phantom}, and particulate gel mechanics~\cite{bantawa2023Hidden}.
By extending this framework to partitioning, our work provides a unified physical description that connects the mechanical and partitioning properties of polymer networks through their topology.
By using a chemically identical PEG--PEG system, we isolated the steric exclusion arising purely from topological constraints.
This approach extends classical partitioning theories for rigid porous media, such as the Casassa model~\cite{casassa1967Equilibrium}, to flexible polymer networks.
In practical systems, chemical interactions between solutes and networks will introduce additional contributions that can be treated as corrections to this topological baseline.
This framework offers predictive insights into solute behavior in diverse network environments, with broad implications for biological systems~\cite{sugano2010Coexistence}, drug delivery~\cite{li2016Designing}, and polymer separation technologies~\cite{gorbunov1988Fundamentals}.

\begin{acknowledgments}
This work was supported by JST ERATO Grant No.~JPMJER2401, CREST Grant No.~JPMJCR1992, and JST FOREST Program Grant No.~JPMJFR232A.
This work was also supported by the Japan Society for the Promotion of Science (JSPS) through the Grants-in-Aid for 
Early-Career Scientists Grant No.~JP25K18074 to T.Y.,
Scientific Research (B) Grant No.~JP23K22458 and No.~JP25K00966 to N.S.,
and Scientific Research (A) Grant No.~JP21H04688 to T.S.
\end{acknowledgments}

\appendix

\section{Comparison of measured $K_0$ with the conventional Ogston model}
We compare our experimentally determined $K_0$ values with the prediction of the conventional Ogston model $K_0= e^{-\phi_\mathrm{excl}}$, where $\phi_\mathrm{excl}= \phi_\mathrm{net}\left(1+R_g/R_f \right)^2$~\cite{ogston1958Spaces, laurent1964Theory, williams1998Partition, hagel2013Diffusion, schnitzer1988Analysis, kotsmar2012Aqueous, tong1996Partitioning, richbourg2023Solute, johnson1995Diffusion}.
Here, $R_f$ is the cross-sectional radius of the network strands, and the solute gyration radius $R_g$ follows the scaling law $R_g \propto M_\mathrm{sol}^{\nu}$~\cite{kawaguchi1997Aqueous}. 

\begin{figure}[t!]
\centering
\includegraphics[width=\linewidth]{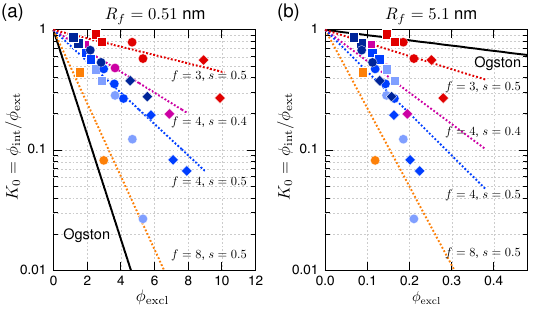}
\caption{Comparison of the measured constant partition ratio $K_0$ with the conventional Ogston model~\cite{ogston1958Spaces, laurent1964Theory}.
The black solid line represents the prediction of the Ogston model as a function of the excluded volume fraction $\phi_{\mathrm{excl}}$, revealing an inconsistency with the measured $K_0$ values (colored symbols).
In (a), the cross-sectional radius of the PEG chain is set to $R_f = 0.51$~nm based on the literature~\cite{hagel2013Diffusion}, while in (b), it is set to $R_f = 5.1$~nm (ten times the literature value) to evaluate the model sensitivity in calculating $\phi_{\mathrm{excl}}$.
}
\label{fig:Ogston}
\end{figure}

Figure~\ref{fig:Ogston} plots $K_0$ against $\phi_\mathrm{excl}$ using (a) $R_f = 0.51$~nm from the literature~\cite{hagel2013Diffusion} and (b) $R_f = 5.1$~nm, a value ten times larger, to test model sensitivity, comparing the measured $K_0$ values in this study (colored symbols) with the prediction of the Ogston model (black solid line).
The measured $K_0$ values vary significantly depending on the network structure (colored dashed lines) and systematically exceed the model predictions in (a).
While increasing $R_f$ in (b) shifts the prediction of the Ogston model toward higher $K_0$ values, the model fails to capture the structural dependence of the measured $K_0$.
These results demonstrate that the Ogston model is inadequate for describing the partitioning of polymer chains into flexible polymer networks, regardless of the choice of $R_f$.

\renewcommand{\theequation}{B\arabic{equation}}
\renewcommand{\theHequation}{B.\arabic{equation}}
\setcounter{equation}{0}

\begin{figure*}[t!]
\centering
\includegraphics[width=\linewidth]{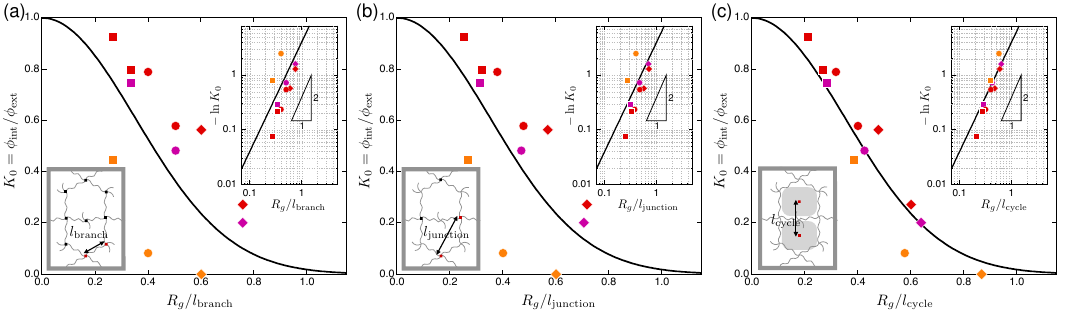}
\caption{The partition ratio $K_0$ for polymer gels with various network topologies (different $f$ and $p$) plotted against alternative characteristic length scales: (a) $R_g/l_\mathrm{branch}$, (b) $R_g/l_\mathrm{junction}$, and (c) $R_g/l_\mathrm{cycle}$.
The $K_0$ values are identical to those in Fig.~\ref{fig:mainresult}.
}
\label{fig:meshsize}
\end{figure*}

\section{Calculation of length scales of the polymer network using the Bethe approximation}
We evaluate the characteristic length scales (mesh sizes) of the polymer network by using the Bethe approximation for star polymer networks~\cite{macosko1976new, miller1976new}: 
the distances between (i) branching points $l_\mathrm{branch}$, (ii) elastically effective junction points $l_\mathrm{junction}$, and (iii) centroids of elastically effective cycles $l_\mathrm{cycle}$.
Consider an $AB$-type polymerization system formed by mixing $f$-functional precursors with terminal functional groups $A$ and $B$ at a molar fraction $s:(1-s)$.
We define $p_A$ ($0 < p_A \leq 1$) and $p_B$ ($0 < p_B \leq 1$) as the fractions of reacted $A$- and $B$-terminal functional groups, respectively.
The overall conversion is $p \equiv s p_A + (1 - s)p_B$, and the stoichiometric balance $s p_A = (1 - s) p_B$ yields $p_A = p/(2s)$ and $p_B = p/[2(1 - s)]$.
Within the Bethe approximation, let $Q_A$ and $Q_B$ be the fractions of $A$- and $B$-terminal functional groups not connected to the infinite network, respectively.
These fractions satisfy the following coupled equations:
\begin{equation}
\left\{ \,
\begin{aligned}
    Q_A & = p_A Q_B^{f - 1} + 1 - p_A,\\
    Q_B & = p_B Q_A^{f - 1} + 1 - p_B.
\end{aligned}
\right.
\label{eq:Q_A}
\end{equation}
By solving Eq.~(\ref{eq:Q_A}) for the nontrivial (unique) solutions within the range $0 \leq Q_A < 1$ and $0 \leq Q_B < 1$, we obtain $Q_A$ and $Q_B$.
Let $X_{A}^{(n)} = \binom{f}{n} Q_A^{f - n} (1 - Q_A)^{n}$ and $X_{B}^{(n)} = \binom{f}{n} Q_B^{f - n} (1 - Q_B)^{n}$ be the fractions of precursors forming $n$-functional elastically effective junction points, where $\binom{f}{n}=f!/[n!(f-n)!]$ is the binomial coefficient.
The numbers of elastically effective chains $\hat{\nu}$ and junction points $\hat{\mu}$ per precursor are given by
\begin{equation}
\begin{aligned}
& \hat{\nu} = \frac{1}{2} \left[ s\sum^{f}_{n=3} {n} X_{A}^{(n)} + (1-s)\sum^{f}_{n=3} {n} X_{B}^{(n)} \right],  \\
& \hat{\mu} = s \sum^{f}_{n=3} X_{A}^{(n)} + (1-s)\sum^{f}_{n=3} X_{B}^{(n)}.
\end{aligned}
\end{equation}
The number of elastically effective cycles per precursor is then $\hat{\xi} = \hat{\nu} - \hat{\mu}$.
Table~\ref{tab:cycle_density} summarizes the calculated values of $\hat{\nu}$, $\hat{\mu}$, and $\hat{\xi}$ for the gels used in this study.

The characteristic length scales are calculated as follows:
$l_\mathrm{branch} = n^{-1/3}$, where $n = c_\mathrm{net,0}N_A/M_\mathrm{pre}$ is the number density of total precursors and $N_A$ is the Avogadro constant.
Similarly, $l_\mathrm{junction} = \mu^{-1/3}$ and $l_\mathrm{cycle} = \xi^{-1/3}$, where $\mu=\hat{\mu} n$ and $\xi=\hat{\xi} n$ are the number densities of elastically effective junction points and cycles, respectively.

\section{Identification of the length scale governing partitioning}
We demonstrate that $l_\mathrm{cycle}$, rather than $l_\mathrm{branch}$ or $l_\mathrm{junction}$, is the critical parameter for observing the universal partitioning law [Eq.~(\ref{eq:K})].
As shown in Table~\ref{tab:cycle_density}, $l_\mathrm{cycle}\,[=(\hat{\xi}n)^{-1/3}]$ varies independently of $l_\mathrm{branch}\,(=n^{-1/3})$ and $l_\mathrm{junction}\,[=(\hat{\mu}n)^{-1/3}]$ when the network topology changes with $f$ and $p$ (see SM, Sec.~S2 for the values of $l_\mathrm{branch}$, $l_\mathrm{junction}$, and $l_\mathrm{cycle}$ for all experimental conditions).
Note that $l_\mathrm{cycle} = l_\mathrm{branch} = l_\mathrm{junction}$ for $f=4$ and $s=0.5$ because $\hat{\xi} = \hat{\mu} = 1$ at $p=1$.
Figure~\ref{fig:meshsize} shows the measured $K_0$ values for various network topologies (purple, red, and orange symbols) as functions of $R_g/l_\mathrm{branch}$, $R_g/l_\mathrm{junction}$, and $R_g/l_\mathrm{cycle}$, respectively.
In (a) and (b), the measured $K_0$ values do not collapse onto a single curve and show significant variation with changes in $f$ and $p$ when using $R_g/l_\mathrm{branch}$ or $R_g/l_\mathrm{junction}$ as the scaling parameter, which is inconsistent with the universal scaling relation [Eq.~(\ref{eq:K})] demonstrated in Fig.~\ref{fig:mainresult} (black solid curve).
In contrast, as shown in (c), the measured $K_0$ values agree well with the universal scaling relation when using $R_g/l_\mathrm{cycle}$.
These findings confirm that $l_\mathrm{cycle}$ is the length scale governing polymer chain partitioning into polymer networks.

\begin{table}[b!]
\caption{Elastically effective network structure parameters per precursor as functions of functionality $f$ and mixing fraction $s$ under the Bethe approximation: chains $\hat{\nu}$, junction points $\hat{\mu}$, and cycles $\hat{\xi}$.}
\label{tab:cycle_density}
\begin{ruledtabular}
\begin{tabular}{c c c c c}
$f$ & $s$ & $\hat{\nu}$ & $\hat{\mu}$ & $\hat{\xi}$\\
\hline
3 & 0.5 & 1.5 & 1 & 0.5\\
4 & 0.5 & 2 & 1 & 1\\
4 & 0.4 & 1.36 & 0.75 & 0.61\\
8 & 0.5 & 4 & 1 & 3\\
\end{tabular}
\end{ruledtabular}
\end{table}

\bibliographystyle{apsrev}

\end{document}


\title{
Supplemental Material for:\\
``Partitioning law of polymer chains into flexible polymer networks''
}

\author{Haruki~Takarai}
\affiliation{Department of Chemistry and Biotechnology, The University of Tokyo, Hongo, Bunkyo-ku, Tokyo 113-8656, Japan}
\author{Takashi~Yasuda}
\affiliation{Faculty of Advanced Life Science, Hokkaido University, Sapporo, Hokkaido 001-0021, Japan}
\author{Naoyuki~Sakumichi}
\affiliation{Department of Chemistry and Biotechnology, The University of Tokyo, Hongo, Bunkyo-ku, Tokyo 113-8656, Japan}
\affiliation{Faculty of Social Informatics, ZEN University, Shinjuku, Zushi, Kanagawa 249-0007, Japan}
\author{Takamasa~Sakai}
\affiliation{Department of Chemistry and Biotechnology, The University of Tokyo, Hongo, Bunkyo-ku, Tokyo 113-8656, Japan}

\date{\today}

\maketitle

\begin{table}[b!]
\caption{Gyration radius $R_g$ and overlap concentrations $c_\mathrm{ext}^{*}$ and $\phi_\mathrm{ext}^{*}$ for linear-PEG solutions at various molar masses, calculated using Eqs.~(\ref{eq:rg}) and (\ref{eq:cstar}).}
\label{tab:r_g}
\begin{ruledtabular}
\begin{tabular}{c c c c}
{$M_\mathrm{sol}$ (kg$/$mol)} & {$R_g$ (nm)} & {$c_\mathrm{ext}^{*}$ (g$/$L)} & {$\phi_\mathrm{ext}^{*}$}\\
\hline
5 & 2.75&74.5&0.062 \\
10 &4.14&43.9&0.037\\
20 &6.22&25.8&0.022\\
\end{tabular}
\end{ruledtabular}
\end{table}

\begin{table}[b!]
\caption{Osmotic pressure $\Pi_\mathrm{ext}$ for linear-PEG solutions at various molar masses and concentrations, calculated using Eq.~(\ref{eq:Piextsol}).}
\label{tab:PiextvsCext}
\begin{ruledtabular}
\begin{tabular}{c c c c c c}
$M_\mathrm{sol}$ & $c_\mathrm{ext}$ & $\Pi_\mathrm{ext}$ & $M_\mathrm{sol}$ & $c_\mathrm{ext}$ & $\Pi_\mathrm{ext}$ \\
\scriptsize{(kg$/$mol)} & \scriptsize{(g$/$L)} & \scriptsize{(kPa)} & \scriptsize{(kg$/$mol)} & \scriptsize{(g$/$L)} & \scriptsize{(kPa)} \\
 \cline{1-3}  \cline{4-6}
5  & 20  & 12.7  & 10  & 90  & 82.2  \\
5  & 40  & 31.5  & 10  & 120 & 143   \\
5  & 60  & 56.9  & 20  & 10  & 1.75  \\
5  & 80  & 89.7  & 20  & 20  & 4.65  \\
5  & 120 & 179   & 20  & 30  & 8.84  \\
10 & 20  & 7.4   & 20  & 40  & 14.5  \\
10 & 30  & 13.1  & 20  & 60  & 30.6  \\
10 & 45  & 24.5  & 20  & 90  & 68.7  \\
10 & 60  & 39.6  & 20  & 120 & 126   \\
\end{tabular}
\end{ruledtabular}
\end{table}

\section{Calculation of Gyration Radius, Overlap Concentration, and Osmotic Pressure of External Linear Polymer Solutions}

Based on previous studies of aqueous linear-PEG solutions~\cite{noda1981thermodynamic,higo1983osmotic,yasuda2023Universality}, we evaluated the gyration radius $R_g$, the overlap concentrations $c_\mathrm{ext}^*$ and $\phi_\mathrm{ext}^*$, and the osmotic pressure $\Pi_{\mathrm{ext}}$ of the external linear-PEG solutions for each molar mass $M_\mathrm{sol}$ used in this study.
The gyration radius $R_g$ is calculated using the following scaling law~\cite{kawaguchi1997Aqueous}:  
\begin{equation}
    R_g = 1.0685 \, M_\mathrm{sol}^{\nu},
    \label{eq:rg}
\end{equation}
where $R_g$ is expressed in nm and $M_\mathrm{sol}$ in kg$/$mol.
This yields the overlap concentration $c_\mathrm{ext}^*$ as
\begin{equation}
    c_\mathrm{ext}^* = \frac{M_\mathrm{sol}}{4\pi^{3/2}R_g^3 N_A \Psi^*},
    \label{eq:cstar}
\end{equation}
where $N_A$ is the Avogadro constant and $\Psi^{*} \approx 0.24$ for linear polymers~\cite{yasuda2023Universality}.
The corresponding overlap volume fraction is $\phi_\mathrm{ext}^{*} = c_\mathrm{ext}^{*}/(\rho_p + c_\mathrm{ext}^{*})$, where $\rho_p$ is the mass density of the polymer defined in the main text.
The calculated values of $R_g$, $c_\mathrm{ext}^*$, and $\phi_\mathrm{ext}^*$ for each $M_\mathrm{sol}$ are summarized in Table~\ref{tab:r_g}.
We evaluated the osmotic pressure of the external polymer solutions $\Pi_{\mathrm{ext}}$ for each $c_\mathrm{ext}$ with the corresponding $M_\mathrm{sol}$ using the universal equation of state for linear polymer solutions in good solvents~\cite{noda1981thermodynamic,higo1983osmotic,yasuda2023Universality},
\begin{equation}
    \frac{\Pi_{\mathrm{ext}} M_\mathrm{sol}}{{c_{\mathrm{ext}} }RT}
    =
    \mathcal{F}\left(\frac{c_{\mathrm{ext}}}{{c_\mathrm{ext}^*}}\right),
    \label{eq:Piextsol}
\end{equation}
where $\mathcal{F}(x)$ is a dimensionless universal function whose explicit form is given in Refs.~\cite{noda1981thermodynamic,higo1983osmotic,yasuda2023Universality}.
The results are summarized in Table~\ref{tab:PiextvsCext}.

\begin{figure*}[t!]
\centering
\includegraphics[width=\linewidth]{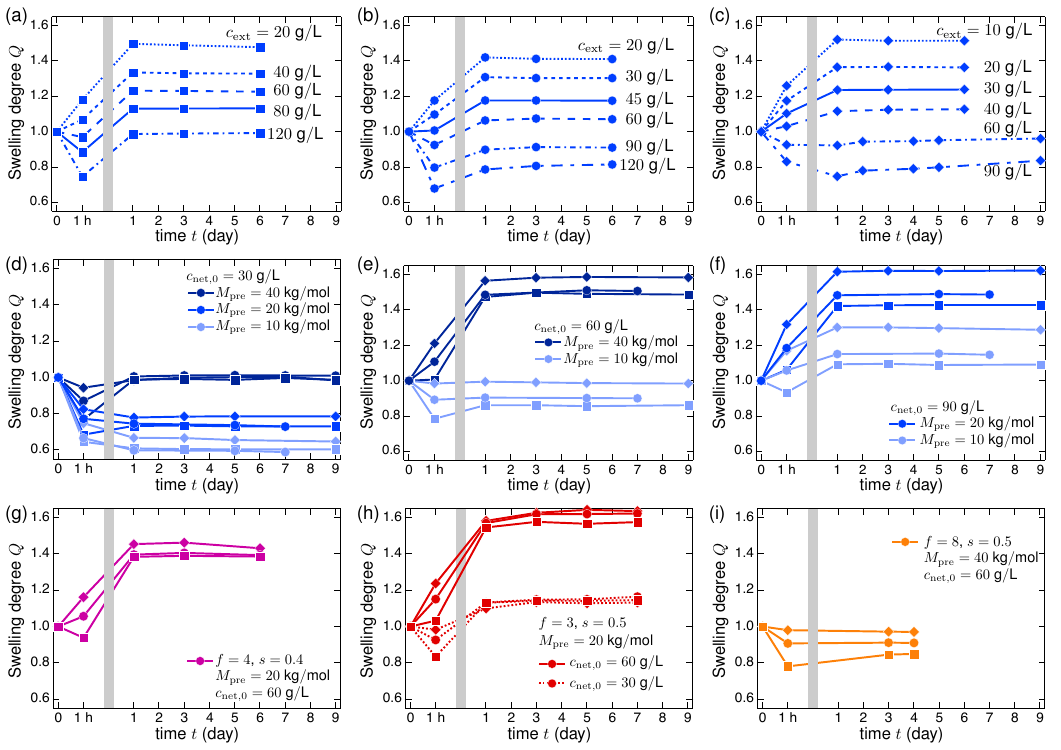}
\caption{Time evolution of the swelling ratio for hydrogels immersed in excess aqueous linear-PEG solutions under varying conditions.
(a)--(c) Hydrogels with $M_\mathrm{pre} = 20$~kg$/$mol and $c_\mathrm{net,0} = 60$~g$/$L in solutions with different polymer solute molar masses: 
(a) $M_{\mathrm{sol}} = 5$~kg$/$mol at $c_{\mathrm{ext}} = 20$--$120$~g$/$L, 
(b) $M_{\mathrm{sol}} = 10$~kg$/$mol at $c_{\mathrm{ext}} = 20$--$120$~g$/$L, 
and (c) $M_{\mathrm{sol}} = 20$~kg$/$mol at $c_{\mathrm{ext}} = 10$--$90$~g$/$L.
(d)--(f) Hydrogels with different precursor concentrations:
(d) $c_\mathrm{net,0} = 30$~g$/$L,
(e) $c_\mathrm{net,0}  = 60$~g$/$L, 
and (f) $c_\mathrm{net,0}  = 90$~g$/$L, each using $M_\mathrm{pre} = 10$, $20$, and $40$~kg$/$mol.
(g)--(i) Hydrogels with different network topologies:
(g) $f = 4$, $s = 0.4$; 
(h) $f = 3$, $s = 0.5$; 
and (i) $f = 8$, $s = 0.5$ at various $M_\mathrm{pre}$ and $c_\mathrm{net,0}$.
All samples in (d)--(i) were immersed in linear-PEG solutions of $M_{\mathrm{sol}} = 5$~kg$/$mol at $c_{\mathrm{ext}} = 80$~g$/$L (squares), $M_{\mathrm{sol}} = 10$~kg$/$mol at $c_{\mathrm{ext}} = 45$~g$/$L (circles), or $M_{\mathrm{sol}} = 20$~kg$/$mol at $c_{\mathrm{ext}} = 30$~g$/$L (diamonds).
}
\label{fig:S-swelling}
\end{figure*}

\section{Swelling Curves and Equilibrium Partitioning Data of Polymer Gels in External Polymer Solutions}

We measured the swelling curves of hydrogels immersed in external linear-PEG solutions because the equilibrium swelling ratio is required to calculate $c_\mathrm{int}$ using Eqs.~(1) and (2) of the main text.
Figure~\ref{fig:S-swelling} summarizes the time evolution of the mass swelling ratio $Q$ for all gel samples examined in this study.
We systematically varied the polymer-network mass concentration ($c_\mathrm{net,0} = 30$, $60$, and $90$~g$/$L), the molar mass of the precursor polymers ($M_\mathrm{pre} = 10$, $20$, and $40$~kg$/$mol), the branching number ($f=3$, $4$, and $8$), and the network connectivity ($p = 2s = 0.8$ and $1.0$).
We determined the equilibrium swelling ratio $Q_\mathrm{eq}$ when the swelling ratio $Q$ remained constant for $2$--$3$ days.
In all cases, except when $c_{\mathrm{ext}}$ significantly exceeded the overlap concentration $c_\mathrm{ext}^{*}$, the gels reached equilibrium within a few days.

Figures~\ref{fig:S-swelling}(a)--(c) show the swelling curves of gels with identical preparation conditions immersed in polymer solutions with different concentrations $c_{\mathrm{ext}}$, demonstrating that $Q_\mathrm{eq}$ varies with the osmotic pressure $\Pi_{\mathrm{ext}}$. 
Figures~\ref{fig:S-swelling}(d)--(f) show the swelling curves of gels with different $c_{\mathrm{net},0}$ and $M_\mathrm{pre}$ immersed in polymer solutions of the same concentration $c_\mathrm{ext}$, indicating that $Q_\mathrm{eq}$ depends on the network structure.
Figures~\ref{fig:S-swelling}(g)--(i) show the swelling curves of gels with different $f$ and $s$, illustrating the effect of network topology on $Q_\mathrm{eq}$.
Tables~\ref{tab:marker_design} and \ref{tab:marker_design_2} summarize the measured $Q_\mathrm{eq}$ and partition ratios $K=\phi_\mathrm{int}/\phi_\mathrm{ext}$ for gel samples with identical network topologies ($f = 4$ and $s = 0.5$) and those with different topologies ($s=0.4$, or $f=3$ and $8$), respectively.
The length scales listed in Tables~\ref{tab:marker_design} and \ref{tab:marker_design_2} were evaluated using the experimentally determined final conversion $p$ of the terminal groups.

\begin{table*}[t!]
\caption{Equilibrium swelling ratios $Q_\mathrm{eq}$ and partition ratios $K$ for polymer chains partitioned into gel networks with $f=4$ and $s=0.5$.}
\label{tab:marker_design}
\begin{ruledtabular}
\begin{tabular*}{\textwidth}{@{\extracolsep{\fill}}cccccccccc}
\multicolumn{2}{c}{\textbf{Gel}} & \multicolumn{2}{c}{\textbf{Solution}} & \multicolumn{3}{c}{\textbf{Results}} & \multicolumn{3}{c}{\textbf{Length scales}} \\
 \cline{1-2} \cline{3-4} \cline{5-7} \cline{8-10}  
$M_\mathrm{pre}$ & $c_\mathrm{net,0}$ & $M_\mathrm{sol}$ & $c_\mathrm{ext}$ & $G_0$ & $Q_\mathrm{eq}$ & $K$ & $l_\mathrm{branch}$ & $l_\mathrm{junction}$ & $l_\mathrm{cycle}$\\
(kg/mol) & (g/L) & (kg/mol) & (g/L) & (kPa) & & & (nm) & (nm) & (nm) \\
\hline
10 & 30 & 5 & 80 & 4.04 & 0.60 & 0.65 & 8.21 & 8.25 & 8.21 \\
10 & 30 & 10 & 45 & 4.04 & 0.60 & 0.29 & 8.21 & 8.25 & 8.21 \\
10 & 30 & 20 & 30 & 4.04 & 0.66 & 0.00 & 8.21 & 8.25 & 8.21 \\
10 & 60 & 5 & 80 & 13.2 & 0.86 & 0.47 & 6.52 & 6.55 & 6.52 \\
10 & 60 & 10 & 45 & 13.2 & 0.90 & 0.12 & 6.52 & 6.55 & 6.52 \\
10 & 60 & 20 & 30 & 13.2 & 0.99 & 0.00 & 6.52 & 6.55 & 6.52 \\
10 & 90 & 5 & 80 & 21.7 & 1.09 & 0.38 & 5.69 & 5.72 & 5.69 \\
10 & 90 & 10 & 45 & 21.7 & 1.15 & 0.03 & 5.69 & 5.72 & 5.69 \\
10 & 90 & 20 & 30 & 21.7 & 1.30 & 0.00 & 5.69 & 5.72 & 5.69 \\
20 & 30 & 5 & 80 & 3.16 & 0.73 & 0.75 & 10.3 & 10.4 & 10.3 \\
20 & 30 & 10 & 45 & 3.16 & 0.74 & 0.48 & 10.3 & 10.4 & 10.3 \\
20 & 30 & 20 & 30 & 3.16 & 0.79 & 0.20 & 10.3 & 10.4 & 10.3 \\
20 & 60 & 5 & 20 & 8.06 & 1.48 & 0.71 & 8.21 & 8.25 & 8.21 \\
20 & 60 & 5 & 40 & 8.06 & 1.33 & 0.68 & 8.21 & 8.25 & 8.21 \\
20 & 60 & 5 & 60 & 8.06 & 1.23 & 0.66 & 8.21 & 8.25 & 8.21 \\
20 & 60 & 5 & 80 & 8.06 & 1.13 & 0.65 & 8.21 & 8.25 & 8.21 \\
20 & 60 & 5 & 120 & 8.06 & 0.99 & 0.63 & 8.21 & 8.25 & 8.21 \\
20 & 60 & 10 & 20 & 8.06 & 1.41 & 0.30 & 8.21 & 8.25 & 8.21 \\
20 & 60 & 10 & 30 & 8.06 & 1.30 & 0.33 & 8.21 & 8.25 & 8.21 \\
20 & 60 & 10 & 45 & 8.06 & 1.18 & 0.36 & 8.21 & 8.25 & 8.21 \\
20 & 60 & 10 & 60 & 8.06 & 1.07 & 0.38 & 8.21 & 8.25 & 8.21 \\
20 & 60 & 10 & 90 & 8.06 & 0.91 & 0.40 & 8.21 & 8.25 & 8.21 \\
20 & 60 & 10 & 120 & 8.06 & 0.81 & 0.43 & 8.21 & 8.25 & 8.21 \\
20 & 60 & 20 & 10 & 8.06 & 1.51 & 0.00 & 8.21 & 8.25 & 8.21 \\
20 & 60 & 20 & 20 & 8.06 & 1.36 & 0.03 & 8.21 & 8.25 & 8.21 \\
20 & 60 & 20 & 30 & 8.06 & 1.24 & 0.08 & 8.21 & 8.25 & 8.21 \\
20 & 60 & 20 & 40 & 8.06 & 1.13 & 0.11 & 8.21 & 8.25 & 8.21 \\
20 & 60 & 20 & 60 & 8.06 & 0.95 & 0.16 & 8.21 & 8.25 & 8.21 \\
20 & 60 & 20 & 90 & 8.06 & 0.80 & 0.23 & 8.21 & 8.25 & 8.21 \\
20 & 60 & 20 & 120 & 8.06 & 0.78 & 0.36 & 8.21 & 8.25 & 8.21 \\
20 & 90 & 5 & 80 & 13.9 & 1.43 & 0.57 & 7.17 & 7.21 & 7.17 \\
20 & 90 & 10 & 45 & 13.9 & 1.49 & 0.27 & 7.17 & 7.21 & 7.17 \\
20 & 90 & 20 & 30 & 13.9 & 1.62 & 0.07 & 7.17 & 7.21 & 7.17 \\
40 & 30 & 5 & 80 & 1.71 & 0.99 & 0.87 & 13.0 & 13.1 & 13.0 \\
40 & 30 & 10 & 45 & 1.71 & 1.01 & 0.68 & 13.0 & 13.1 & 13.0 \\
40 & 30 & 20 & 30 & 1.71 & 1.00 & 0.38 & 13.0 & 13.1 & 13.0 \\
40 & 60 & 5 & 80 & 4.87 & 1.49 & 0.78 & 10.3 & 10.4 & 10.3 \\
40 & 60 & 10 & 45 & 4.87 & 1.51 & 0.54 & 10.3 & 10.4 & 10.3 \\
40 & 60 & 20 & 30 & 4.87 & 1.59 & 0.28 & 10.3 & 10.4 & 10.3 \\
\end{tabular*}
\end{ruledtabular}
\end{table*}


\begin{table*}[t!]
\caption{Equilibrium swelling ratios $Q_\mathrm{eq}$ and partition ratios $K$ for polymer chains partitioned into gels with different network topologies.}
\label{tab:marker_design_2}
\begin{ruledtabular}
\begin{tabular*}{\textwidth}{@{\extracolsep{\fill}}cccccccccc}
\multicolumn{2}{c}{\textbf{Gel}} & \multicolumn{2}{c}{\textbf{Solution}} & \multicolumn{3}{c}{\textbf{Results}}& \multicolumn{3}{c}{\textbf{Length scales}} \\
 \cline{1-2} \cline{3-4} \cline{5-7} \cline{8-10}  
$M_\mathrm{pre}$ & $c_\mathrm{net,0}$ & $M_\mathrm{sol}$ & $c_{\mathrm{ext}}$ & $G_0$ & $Q_\mathrm{eq}$ & $K$ & $l_\mathrm{branch}$ & $l_\mathrm{junction}$ & $l_\mathrm{cycle}$ \\
(kg/mol) & (g/L) & (kg/mol) & (g/L) & (kPa) & & & (nm) & (nm) & (nm) \\
\hline \multicolumn{10}{c}{$f=4$, $s=0.4$} \\ \hline
20 & 60 & 5 & 80 & 5.35 & 1.39 & 0.75 & 8.21 & 8.82 & 9.74 \\
20 & 60 & 10 & 45 & 5.35 & 1.41 & 0.48 & 8.21 & 8.82 & 9.74 \\
20 & 60 & 20 & 30 & 5.35 & 1.46 & 0.20 & 8.21 & 8.82 & 9.74 \\
\hline \multicolumn{10}{c}{$f=3$, $s=0.5$} \\ \hline
20 & 30 & 5 & 80 & 3.16 & 1.14 & 0.93 & 10.3 & 10.9 & 13.0 \\
20 & 30 & 10 & 45 & 3.16 & 1.16 & 0.79 & 10.3 & 10.9 & 13.0 \\
20 & 30 & 20 & 30 & 3.16 & 1.13 & 0.56 & 10.3 & 10.9 & 13.0 \\
20 & 60 & 5 & 80 & 3.86 & 1.58 & 0.80 & 8.21 & 8.67 & 10.3 \\
20 & 60 & 10 & 45 & 3.86 & 1.63 & 0.58 & 8.21 & 8.67 & 10.3 \\
20 & 60 & 20 & 30 & 3.86 & 1.64 & 0.27 & 8.21 & 8.67 & 10.3 \\
\hline \multicolumn{10}{c}{$f=8$, $s=0.5$} \\ \hline
40 & 60 & 5 & 80 & 10.2 & 0.85 & 0.45 & 10.3 & 10.3 & 7.17 \\
40 & 60 & 10 & 45 & 10.2 & 0.91 & 0.08 & 10.3 & 10.3 & 7.17 \\
40 & 60 & 20 & 30 & 10.2 & 0.97 & 0.00 & 10.3 & 10.3 & 7.17 \\
\end{tabular*}
\end{ruledtabular}
\end{table*}


\begin{figure}[t!]
\centering
\includegraphics[width=0.98\linewidth]{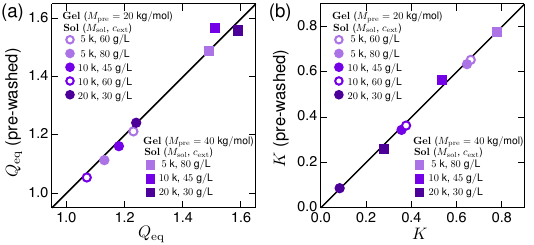}
\caption{Comparison of (a) equilibrium swelling ratio $Q_\mathrm{eq}$ and (b) partition ratio $K$ between polymer gels that were pre-washed to remove buffer salts (vertical axis) and those that were not washed and thus contained the buffer salts initially (horizontal axis).
Gels were prepared at $c_\mathrm{net,0}=60~\mathrm{g/L}$ with $M_{\mathrm{pre}}=20$ (circles) or $40$~kg$/$mol (squares).
The color gradient from light to dark purple indicates increasing solute molar masses: $M_{\mathrm{sol}} = 5$, $10$, and $20$~kg$/$mol.
Across all investigated conditions, both $Q_{\mathrm{eq}}$ and $K$ are consistent between the two protocols.
}
\label{fig:ion}
\end{figure}

\section{Negligible effect of buffer salts on swelling and partitioning}

To verify that buffer salts do not play a significant role in the equilibrium properties in the investigated regime, we compared the equilibrium swelling ratio $Q_\mathrm{eq}$ and the partition ratio $K$ of polymer gels that were pre-washed in excess deionized water to remove buffer salts (computed from the weights $W_0$ and $W_\mathrm{eq}$ listed in Table~\ref{tab:Dryweight} in the next section) with those that were not washed and thus contained the buffer salts initially (data in Table~\ref{tab:marker_design}).
Figures~\ref{fig:ion}(a) and (b) plot $Q_{\mathrm{eq}}$ and $K$ for the pre-washed gels against those for the unwashed gels, respectively.
Across all investigated conditions, $Q_{\mathrm{eq}}$ and $K$ were consistent between the two protocols, indicating that buffer salts rapidly diffuse into the pure water outside the gel and become extremely dilute, and thus do not contribute to the measurable interactions with PEG.
This result demonstrates that buffer salts do not measurably influence swelling or partitioning.
These observations confirm that the buffer salts can be neglected, allowing the system to be described entirely by the network, the polymer solute, and the solvent.


\begin{figure}[t!]
\centering
\includegraphics[width=\linewidth]{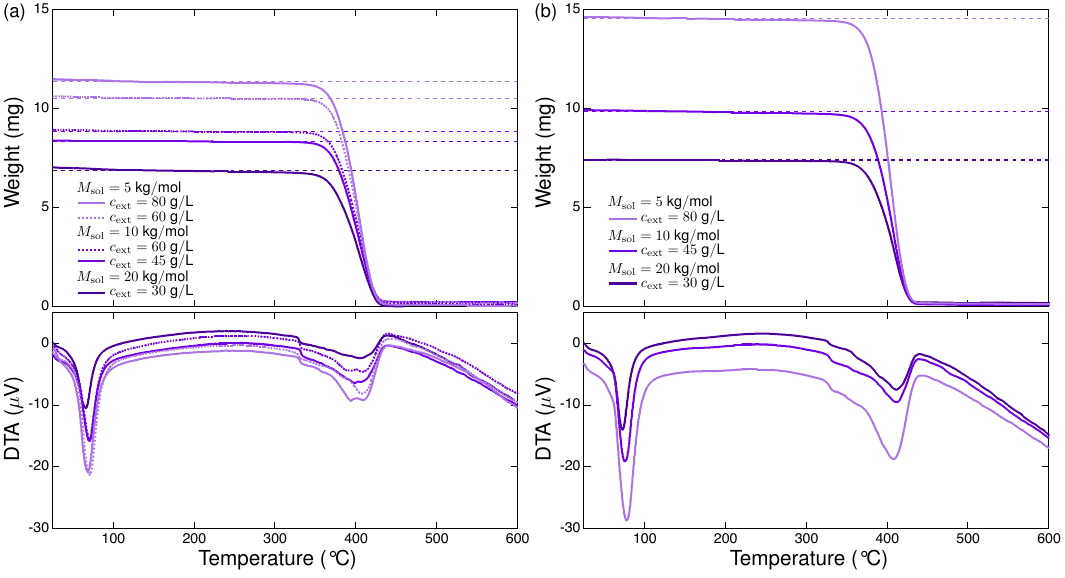}
\caption{
Thermal profiles of hydrogels containing partitioned polymer chains, vacuum-dried at $T = 333$~K, obtained using thermogravimetric/differential thermal analysis (TG/DTA). 
Both hydrogels have a network concentration $c_{\mathrm{net,0}} = 60$~g$/$L with precursor molar masses of (a) $M_\mathrm{pre} = 20$~kg$/$mol and (b) $M_\mathrm{pre} = 40$~kg$/$mol.
The color gradient from light to dark purple indicates increasing solute molar masses: $M_{\mathrm{sol}} = 5$, $10$, and $20$~kg$/$mol.
The horizontal dashed lines indicate the residual weight at $150\,{}^\circ\mathrm{C}$, corresponding to the total PEG content in the hydrogel; the weight curves remain on these lines up to $T \approx 400\,{}^\circ\mathrm{C}$.
}
\label{fig:TGA}
\end{figure}

\begin{table}[t!]
\caption{Measured weights of the pre-washed gels in the as-prepared state $W_0$, equilibrium swollen state $W_\mathrm{eq}$, and dried state $W_\mathrm{dry}$.}
\label{tab:Dryweight}
\begin{ruledtabular}
\begin{tabular}{c c c c c c c}
{$M_{\mathrm{pre}}$} & {$c_{\mathrm{net,0}}$} & {$M_{\mathrm{sol}}$} & {$c_{\mathrm{ext}}$} & {$W_0$} & {$W_\mathrm{eq}$} & {$W_\mathrm{dry}$} \\
\scriptsize{(kg$/$mol)}&\scriptsize{(g$/$L)}&\scriptsize{(kg$/$mol)}&\scriptsize{(g$/$L)}&\scriptsize{(mg)}&\scriptsize{(mg)}&\scriptsize{(mg)}\\
\hline
20 & 60 & 5  & 60 & 111.6 & 135.3& 10.28\\
20 & 60 & 5  & 80 & 110.4 & 122.9& 11.25\\
20 & 60 & 10 & 45 & 112.5 & 130.7 &8.08\\
20 & 60 & 10 & 60 & 111.2 & 117.2 &8.69\\
20 & 60 & 20 & 30 & 110.7 & 137.5 &6.83\\
40 & 60 & 5  & 80 & 112.5 & 167.4 &14.32\\
40 & 60 & 10 & 45 & 110.9 & 174.0 &9.68\\
40 & 60 & 20 & 30 & 112.7 & 175.7 &7.15\\
\end{tabular}
\end{ruledtabular}
\end{table}

\section{Thermogravimetric Measurement of Dry Weight of Hydrogels Containing Partitioned Polymer Chains}

We measured the dry weight $W_{\mathrm{dry}}$ of the hydrogels containing partitioned polymer chains using thermogravimetric analysis (TGA) to directly quantify $c_\mathrm{int}$.
Figure~\ref{fig:TGA} summarizes the results of thermogravimetric and differential thermal analyses of the samples after vacuum drying at $T = 333$~K.
The gel weight remained approximately constant in the temperature range $T=150$--$400 \,{}^\circ\mathrm{C}$, indicating that the water in the hydrogels had been completely removed, and decreased to approximately zero for $T > 450\,{}^\circ\mathrm{C}$, at which PEG was completely decomposed.
We determined $W_\mathrm{dry}$ for each gel sample as the residual weight at $T = 150\,{}^\circ\mathrm{C}$ (horizontal dashed lines in Fig.~\ref{fig:TGA}); the near-constancy of the weight up to $T \approx 400\,{}^\circ\mathrm{C}$ confirms that this value is insensitive to the choice of temperature within this range.
The results are summarized in Table~\ref{tab:Dryweight}.

\newpage

\section{Determination of Shear Modulus for Evaluating Elastic Contribution in Polymer Gels}

\begin{figure}[t!]
\centering
\includegraphics[width=0.6\linewidth]{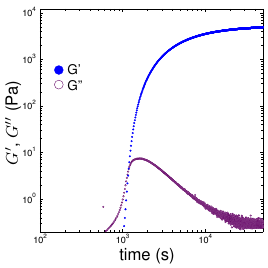}
\caption{Typical result of the time evolution of the storage modulus $G'$ and loss modulus $G''$ for a sample with $c_\mathrm{net,0} = 60$~g$/$L, $M_\mathrm{pre} = 40$~kg$/$mol, $s=0.5$, and $f=4$~\cite{yoshikawa2021Negative}.
The measurement was performed at a constant frequency of $1$~Hz using a double-cylinder rheometer.
The storage modulus $G'$ increases during the reaction and reaches a stable plateau.
This plateau $G'$ value defines the shear modulus $G_0$ in the as-prepared state used in the calculation of $\Pi_{\mathrm{el}}$.
}
\label{fig:g0}
\end{figure}

To determine the elastic contribution of the gel network, $\Pi_{\mathrm{el}}$, which follows the scaling relationship $\Pi_{\mathrm{el}} = -G_0(c_{\mathrm{net}}/c_{\mathrm{net,0}})^{1/3}$~\cite{james1949simple, horkay2000osmotic}, we utilized rheological data from our previous study~\cite{yoshikawa2021Negative} to determine the shear modulus $G_0$ of polymer gels in the as-prepared state for each network condition.
Here, we assume that the presence of partitioned polymer chains only slightly alters the shear modulus of the network, as established by Horkay and Zrinyi~\cite{horkay1986Studies}.
Figure~\ref{fig:g0} shows a typical result of the time evolution of the storage modulus $G'$ and loss modulus $G''$ in a rheological measurement.
The gelation process was measured at constant frequency, strain, and temperature ($25\,{}^\circ\mathrm{C}$) using a double-cylinder rheometer.
The storage modulus $G'$ increases as network formation proceeds and reaches a steady plateau.
We define this plateau value as the shear modulus $G_0$ in the as-prepared state.
The values of $G_0$ measured for each network are summarized in Tables~\ref{tab:marker_design} and \ref{tab:marker_design_2}.

\bibliographystyle{apsrev}